\documentclass[]{raa}            % referee version: for submission
\usepackage{graphicx,times}
\usepackage{natbib}

\providecommand{\bm}[1]{\mbox{\boldmath$#1$\unboldmath}}

\begin{document}

   \title{The Color Excesses of Type Ia Supernovae from Single-Degenerate Channel Model
% $^*$
%\footnotetext{\small $*$ Supported by the National Natural Science Foundation of China.}
}

 \volnopage{ {\bf 2009} Vol.\ {\bf 9} No. {\bf XX}, 000--000}
   \setcounter{page}{1}

   \author{X.-C. Meng
      \inst{1}
   \and X.-F. Chen
      \inst{2}
   \and Z.-W. Han
      \inst{2}
   \and W.-M. Yang
     \inst{1}
   }

   \institute{Department of Physics and Chemistry, Henan Polytechnic
University, Jiaozuo, 454000, China; {\it conson859@msn.com}\\
%        \and
%             Full institute address for the second author
        \and
             National Astronomical Observatories/Yunnan
Observatory, the Chinese Academy of Sciences, Kunming, 650011,
China
%\vs \no
%   {\small Received [year] [month] [day]; accepted [year] [month] [day] }
}

\abstract{Single Degenerate model is the most widely accepted progenitor model of type Ia supernovae (SNe Ia), in which
   a carbon-oxygen white dwarf (CO WD) accretes hydrogen-rich
   material from a main sequence or a slightly evolved star (WD +MS) to increase its mass, and explodes when
   its mass approaches the Chandrasekhar mass limit. During the
   mass transfer phase between the two components, an optically thick
   wind may occur and the material lost as the wind
   may exist as circumstellar material (CSM). Searching the CSM around
   progenitor star is helpful to discriminate different progenitor models
   of SNe Ia. Meanwhile, the CSM is a source of
   color excess.The purpose of this paper is to study the color excess produced from the single-degenerate
   progenitor model with optically thick wind, and reproduce
   the distribution of color excesses of SNe Ia. Meng et al. (2009) systemically carried out binary evolution calculation of
   the WD +MS systems for various metallicities and showed the parameters of the systems before
   Roche lobe overflow and at the moment of supernova explosion in Meng \& Yang (2009).
   With the results of Meng et al. (2009), we calculate the color
   excesses of SNe Ia at maximum light via a simple analytic method.We reproduces the distribution of color excesses of SNe Ia
   by our binary population synthesis approach if
   the velocity of the optically thick wind is taken to be of order of magnitude of 10 km s$^{\rm -1}$. However,
   if the wind velocity is larger than 100 km s$^{\rm -1}$, the reproduction is bad.
 \keywords{Stars: white dwarfs - stars: supernova: general } }

   \authorrunning{X.-C. Meng, X.-F. Chen, Z.-W. Han \& W.-M. Yang}            %author_head in even pages
   \titlerunning{The Color Excesses of Type Ia Supernovae from Single-Degenerate Channel Model}  % title_head in odd pages
   \maketitle

%% The author head (on even pages) and the title head (on odd pages) will be
%% automatically extracted from \author{} and \title{}. Whenever the title is too long,
%% you will be asked to supply a shorter one by inserting either \authorrunning{} or
%% \titlerunning{} before \maketitle. Anyway, you can specify your own heads.
%%
%%
%% Note: In the following text body of your manuscript, please note several differences from
%%       other major journals:
%% (1) \subsection{Please Capitalize the First Letter of Each Notional Word in Subsection Title}
%% (2) Please Capitalize the First Letter of Each Notional Word in all tables' captions

%
%________________________________________________ sections below
%
\section{INTRODUCTION}           %% first-level sections will be auto-capitalized
\label{sect:1}

Although type Ia supernovae (SNe Ia) showed their importance in
determining cosmological parameters, e.g. $\Omega_{\rm M}$ and
$\Omega_{\Lambda}$ (\citealt{RIE98}; \citealt{PER99}), the
progenitor systems of SNe Ia have not been confidently identified
yet (\citealt{HN00}; \citealt{LEI00}). It is widely believed that
a SN Ia is from thermonuclear runaway of a carbon-oxygen white
dwarf (CO WD) in a binary system. The CO WD accretes material from
its companion to increase its mass. When its mass reaches its
maximum stable mass, it explodes as a thermonuclear runaway and
almost half of the WD mass is converted into radioactive nickel-56
(\citealt{BRA04}). Two basic scenarios have been discussed over
the last three decades. One is a single degenerate (SD) model,
which is widely accepted (\citealt{WI73}). In this model, a CO WD
increases its mass by accreting hydrogen- or helium-rich matter
from its companion, and explodes when its mass approaches the
Chandrasekhar mass limit. The companion may be a main-sequence
star (WD+MS) or a red-giant star (WD+RG) (\citealt{YUN95};
\citealt{LI97}; \citealt{HAC99a}, \citealt{HAC99b};
\citealt{NOM99, NOM03}; \citealt{LAN00}; \citealt{HAN04};
\citealt{CHENWC07}; \citealt{HAN08}; \citealt{MENGXC09a};
\citealt{LGL09}). An alternative is the theoretically less favored
double degenerate (DD) model (\citealt{IT84}; \citealt{WEB84}), in
which a system consisting of two CO WDs loses orbital angular
momentum by gravitational wave radiation and merges finally. The
merger may explode if the total mass of the system exceeds the
Chandrasekhar mass limit (see the reviews by \citealt{HN00} and
\citealt{LEI00}). In theory, a large amount of circumstellar
materials (CSM) may be produced via an optically thick wind for
the SD model (\citealt{HAC96}), while there is no CSM around DD
systems. Then, a basic method to distinguish the two progenitor
models is to find the CSM around progenitor systems.

Evidence for CSM was first found in SN2002ic (\citealt{HAMUY03}),
which has shown extremely pronounced hydrogen emission lines that
have been interpreted as a sign of strong interaction between
supernova ejecta and CSM. The discovery of SN2002ic may uphold the
SD model (\citealt{HAN06}). Recently, the evidence for CSM was
found in a normal SN Ia (SN 2006X) defined by \citet{BRA93} and
the CSM is proposed to be from a wind from a red-giant companion
(\citealt{PATAT07a}). The CSM may play a key role to solve the
problem of the low value of reddening ratio of external galaxy
(\citealt{WANGLF05}), which is very important for precision
cosmology (\citealt{WANGXF07}).

If a SN Ia is surrounded by a large amount of CSM, its color
observed should be redder than its intrinsic color, which results
in a color excess, $E(B-V)$. \citet{REINDL05} showed the color
excesses of more than one hundred SNe Ia at maximum light, which
suggests a mission to check which progenitor model of SNe Ia can
explain the distribution of the color excesses. Recently,
\citet{MENGXC09a} performed binary stellar evolution calculations
for more than 25,000 close WD binary systems with various
metallicities, and present all the parameters of the systems for
SNe Ia before the Roche lobe overflow (RLOF) and at the moment of
supernova explosion in a following paper (\citealt{MENGXC09b}). In
their works, the prescription of \citet{HAC99a} for the accretion
efficiency of hydrogen-rich material was adopted by assuming an
optically thick wind (\citealt{HAC96}), and then their works
provide a possibility to check whether the SD model with optically
thick wind can reproduce the distribution of color excesses of SNe
Ia obtained from observation or not. The purpose of this paper is
to check the possibility, and this work is based on the results
from \citet{MENGXC09a}.

In section \ref{sect:2}, we describe our model. We show the
results in section \ref{sect:3} and give discussions and
conclusions in sections \ref{sect:4} and \ref{sect:5}.

% Authors can give a citation as `Michel et al. 1992'.
% You may also use \cite, \citep and \citet for citation, and use Table~1
% or Figure~1 and so forth. Using \ref and \label for cross-references of
% Tables/Figures is a good way in adjusting/adding/removing text, tables or
% figures.

\section{Model and Physics Inputs}
\label{sect:2}

   \begin{figure}[h!!!]
   \centering
   \includegraphics[width=9.0cm, angle=270]{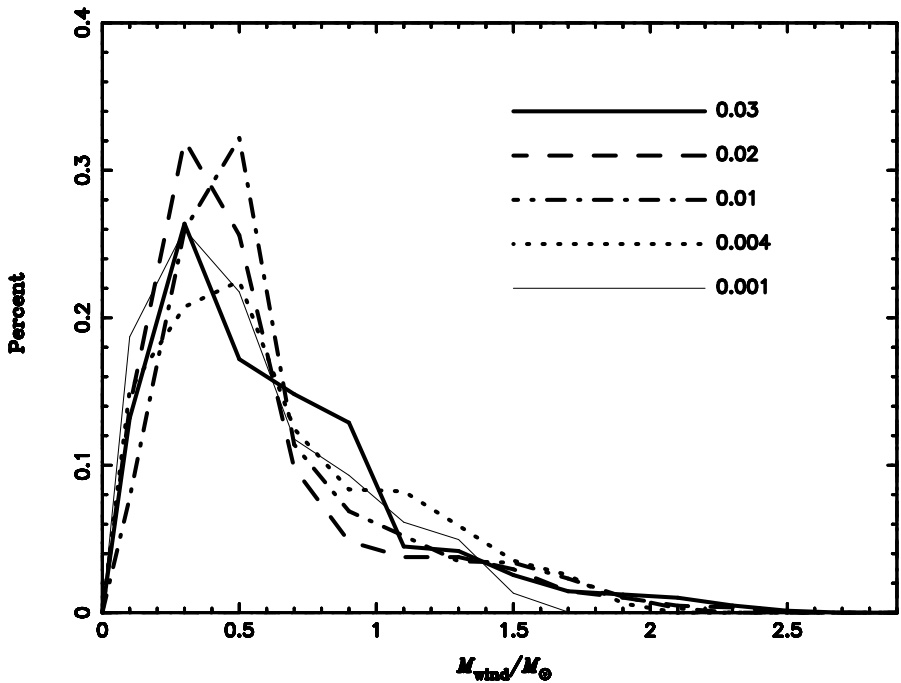}

%   \begin{minipage}[]{85mm}

   \caption{The mass distribution of hydrogen-rich material lost as optically thick
wind for different metallicities.}
%   \end{minipage}
   \label{lostm}
   \end{figure}

\subsection{the mass distribution of lost hydrogen-rich material}
\label{sect:2.1}

As described in section \ref{sect:1}, the color of a supernova is
reddened by the CSM surrounding the supernova. We first check
whether there exists enough CSM resulting from optically thick
wind. The following is a simple description about how to obtain
the CSM. As described in \citet{MENGXC09a}, in a WD + MS system,
the companion fills its Roche lobe at MS or during HG, and
transfers material onto the WD. If the mass-transfer rate,
$|\dot{M}_{\rm 2}|$, exceeds a critical value, $\dot{M}_{\rm cr}$,
we assume that the accreted hydrogen steadily burns on the surface
of WD and that the hydrogen-rich material is converted into helium
at the rate of $\dot{M}_{\rm cr}$. The unprocessed matter is
assumed to be lost from the system as an optically thick wind at a
rate of $\dot{M}_{\rm wind}=|\dot{M}_{\rm 2}|-\dot{M}_{\rm cr}$
(\citealt{HAC96}). Adopting the prescription of \citet{HAC99a} on
WDs accreting hydrogen-rich material from their companions,
\citet{MENGXC09a} obtained the initial parameters of WD + MS
systems for SNe Ia, and at the same time, the final parameters at
the moment of SN Ia explosion, such as $M_{\rm 2}^{\rm SN}$. In
this paper, incorporating the binary evolution results in
\citet{MENGXC09a} into the rapid binary evolution code developed
by Hurley et al. (2000, 2002), we carry out a series of binary
population synthesis (BPS) studies for various Z. In each BPS
study, $10^{\rm 7}$ binaries are generated by Monte Carlo
simulation and a circular orbit is assumed for all binaries. The
basic parameters for the simulations are same to those in
\citet{MENGXC09a}. \textbf{It can be shown that a WD+MS system may
originate from three possible evolution channels, namely, the He
star channel, the EAGB channel and the TPAGB channel (see
\citealt{MENGXC09a} for details)}. We assume that a SN Ia is
produced if the initial parameters of a WD + MS system, e.g.
initial orbital period $P_{\rm orb}^{\rm i}$ and initial secondary
mass $M_{\rm 2}^{\rm i}$, locate in the appropriate regions of the
parameters for SNe Ia at the onset of RLOF. We can obtain the
companion mass at the moment of explosion by interpolation in the
three-dimensional grid ($M_{\rm WD}^{\rm i}, M_{\rm 2}^{\rm i},
\log P^{\rm i}$) of the more than 25,000 close WD binary system
calculated in \citet{MENGXC09a}. In Fig. \ref{lostm}, we show the
mass distribution of hydrogen-rich material lost as the optically
thick wind for various metallicities, $M_{\rm wind}=(M_{\rm
2}^{\rm i}+M_{\rm WD}^{\rm i})-(M_{\rm 2}^{\rm SN}+M_{\rm WD}^{\rm
SN})$, where superscript i and SN represent the initial and final
values for white dwarf and secondary, respectively, and $M_{\rm
WD}^{\rm SN}$ is assumed to be $1.378 M_{\odot}$. We can see from
the figure that the distribution of the lost mass peaks at about
0.3 $M_{\odot}$ and has a high-mass tail. The amount of the lost
material may be as large as 2.5 $M_{\odot}$, which should
contribute to the color excess of SNe Ia.

In Fig. \ref{lostm}, we can see that there does not seems to be a
systemic trend with metallicity. Actually, the influence of
metallicity on the $M_{\rm wind}$ is complicated. $M_{\rm wind}$
is mainly determined by $M_{\rm 2}^{\rm i}$ and $M_{\rm WD}^{\rm
i}$. The two parameters are both systemically affected by
metallicity, but the tendency is reversed, i.e. the peak of the
distribution of companion move to higher mass with metallicity,
while the peak for WD mass to lower mass (see Figs. 9 and 10 in
\citealt{MENGXC09a}). In addition, the metallicity also affect the
mass-transfer rate between WD and its companion, and then $M_{\rm
2}^{\rm SN}$ (\citealt{LAN00}). The complicated influence of
metallicity on the distribution of $M_{\rm wind}$ results in a
non- systemic trend of the distribution with metallicity. We also
noticed that the percentage of high $M_{\rm wind}$, e.g. $M_{\rm
wind}>2.0M_{\odot}$, increases with metalliticity. High $M_{\rm
wind}$ is mainly determined by the \textbf{upper} boundary of the
companion mass, which moves to higher mass with metallicity (see
Fig. 4 in \citealt{MENGXC09a}). So, a binary system producing SN
Ia with high metallicity may loss more hydrogen-rich material by
optically thick wind, and then the high-metallicity model in Fig.
\ref{lostm} shows a higher percentage of high $M_{\rm wind}$.

\subsection{model}
\label{sect:2.2}

\begin{figure}
   \centering
%   \vspace{2mm}
%   \begin{center}
   %%%\includegraphics{empty.eps}
   %%%\includegraphics{empty.eps}
   \includegraphics[width=90mm,angle=270.0]{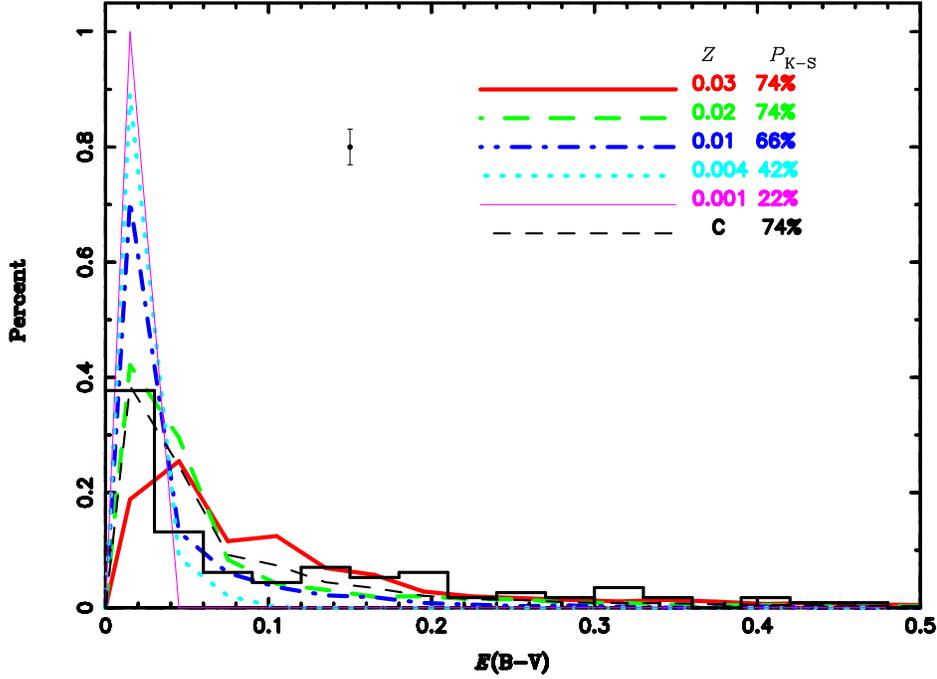}
%   \hspace{3mm}
%   \psfig {figure=mix.ps,width=80mm,height=60mm,angle=270.0}
%   \parbox{180mm}{{\vspace{2mm} }}
   \caption{The distribution of color excesses of SNe Ia at maximum light for various metallicties,
   where wind velocity is assumed to be 10 km s$^{\rm -1}$. The
   solid histogram is from observation, and the bar represents its maximum error (\citealt{REINDL05}).
   The black dashed line is the sum of those of Z=0.01, 0.02 and 0.03, where
   the weights for the three components are 20\%, 40\% and 40\% respectively. Via K-S test, the
   percentages indicated in the figure
   show the confident level that the distributions of color excess from theory and observation are indistinguishable. }
              \label{ebvzz}%
    \end{figure}

\citet{BOHLIN78} found
\begin{equation}
E(B-V)/n_{\rm H}=1.72\times10^{\rm -22}{\rm mag}\hspace{0.1cm}
{\rm cm^{\rm 2}}, \label{eq:ebv}
  \end{equation}
for the Galaxy, where $n_{\rm H}$ is total hydrogen column density
and this relation linearly depends on metallicity
(\citealt{DRAINE03}). The linear relation can be fitted by
 \begin{equation}
E(B-V)/n_{\rm H}=(17.4\times Z/Z_{\odot}-0.454)\times10^{\rm -23}
{\rm mag}\hspace{0.1cm} {\rm cm^{\rm 2}},
  \end{equation}
where the fitting data are from \citet{BOHLIN78},
\citet{KOORNEEF82}, \citet{FITZPATRICK85} and \citet{MARTIN89}. In
the following, we describe how to obtain the total hydrogen column
density, $n_{\rm H}$.

\textbf{While the} CSM around the progenitor of a SN Ia may be
asymmetric, \textbf{for} simplicity, we assume that the mass loss
of optically thick wind is spherically symmetric and the mass-loss
rate is constant during the whole mass-transfer phase. Then,
hydrogen number density is only a function of the distance of wind
material from the progenitor star, i.e.
\begin{equation}
 n(r)=ar^{\rm -2},
  \end{equation}
where $a$ is a coefficient determined by
 \begin{equation}
 \int_{r_{\rm 0}}^{r_{\rm m}}ar^{\rm -2}\cdot4\pi r^{\rm 2}dr=N_{\rm
 H},
  \end{equation}
where $r_{\rm 0}$ is the radius of progenitor, $r_{\rm m}$ is the
maximum distance which hydrogen-rich material can arrive as
optically thick wind at the moment of SN Ia explosion, and $N_{\rm
 H}$ is the total number of hydrogen atom. The
total hydrogen column density at maximum light is calculated from
 \begin{equation}
 \int_{r_{\rm 1}}^{r_{\rm m}}ar^{\rm -2}dr=n_{\rm H},
  \end{equation}
where $r_{\rm 1}$ is the distance of explosion ejecta from
explosion center at maximum light. From equations (4) and (5), we
can obtain
 \begin{equation}
n_{\rm H}=\frac{N_{\rm H}}{4\pi r_{\rm m}r_{\rm 1}}. \label{eq:nh}
  \end{equation}
$N_{\rm H}$ can be obtained from
 \begin{equation}
 N_{\rm H}=\frac{XM_{\rm wind}}{m_{\rm H}},
  \end{equation}
where $M_{\rm wind}$ is the total mass lost as optically thick
wind, $m_{\rm H}$ is the mass of a hydrogen atom and $X$ is the
mass fraction of hydrogen and is set to be
\begin{equation}
 X=0.76-3.0Z,
  \end{equation}
(\citealt{POL98}). $r_{\rm m}$ can be obtained from
 \begin{equation}
 r_{\rm m}=V_{\rm wind}t_{\rm d},
  \end{equation}
where $V_{\rm wind}$ is the velocity of the optically thick wind,
and $t_{\rm d}$ is the delayed time from the onset of mass
transfer to the moment of SN Ia explosion. From the results of
\citet{MENGXC09a}, $t_{\rm d}$ can be approximated by
 \begin{equation}
 \log(t_{\rm d}/{\rm yr})=-\frac{2}{3}M_{\rm 2}^{\rm i}+7.8,
  \end{equation}
where $M_{\rm 2}^{\rm i}$ is the initial mass of the secondary
(the mass donor unit in solar mass) in a WD + MS system. $t_{\rm
d}$ from the equation is a rough estimation of the mean value for
a certain $M_{\rm 2}^{\rm i}$ and has an error of about 50\%.
$r_{\rm 1}$, the distance of explosion ejecta from explosion
center at maximum light, can be obtained by the product of the
velocity of supernova ejecta and the rise time of light curve of
SNe Ia. We simply assume that the velocity of ejecta is 10000 km
s$^{\rm -1}$ (\citealt{GAM03}) and the rise time of light curve of
SNe Ia is 20 day (\citealt{CONLEY06}; \citealt{STROVINK07}). The
ejecta velocity adopted here corresponds to a typical photospheric
velocity (\citealt{WANGLF03}) and might be lower than terminal
ejecta velocity (\citealt{WANGLF06}). However, the uncertainty
resulting from the ejecta velocity is moderate and accepted. We
will discuss its influence in section \ref{sect:4}. The rise time
does not significantly affect the final results.

\citet{MENGXC09a} performed binary stellar evolution calculations
for more than 25,000 close WD binary systems with various
metallicities and \citet{MENGXC09b} presented the distribution of
all the parameters for these close systems before the RLOF and at
the moment of SN Ia explosion. Incorporating their results into
the binary population synthesis code of \citet{HUR00, HUR02}, we
obtain the distribution of the wind mass, $M_{\rm wind}$ (see Fig.
\ref{lostm}) and the color excess via equations \ref{eq:ebv} and
\ref{eq:nh}. The basic parameters for Monte Carlo simulations are
same to that in \citet{MENGXC09a} when primordial binary samples
are generated. Because the code is valid just for $Z\leq0.03$,
only five metallicities (i.e. Z = 0.03, 0.02, 0.01, 0.004 and
0.001) are examined here.

The greatest uncertainty of our model is from $V_{\rm wind}$.
Here, we assume that $V_{\rm wind}=10$ km s$^{\rm -1}$ and we will
discuss it in section \ref{sect:4}.

   \begin{figure}
   \centering
%   \vspace{2mm}
%   \begin{center}
   %%%\includegraphics{empty.eps}
   %%%\includegraphics{empty.eps}
   \includegraphics[width=90mm,angle=270.0]{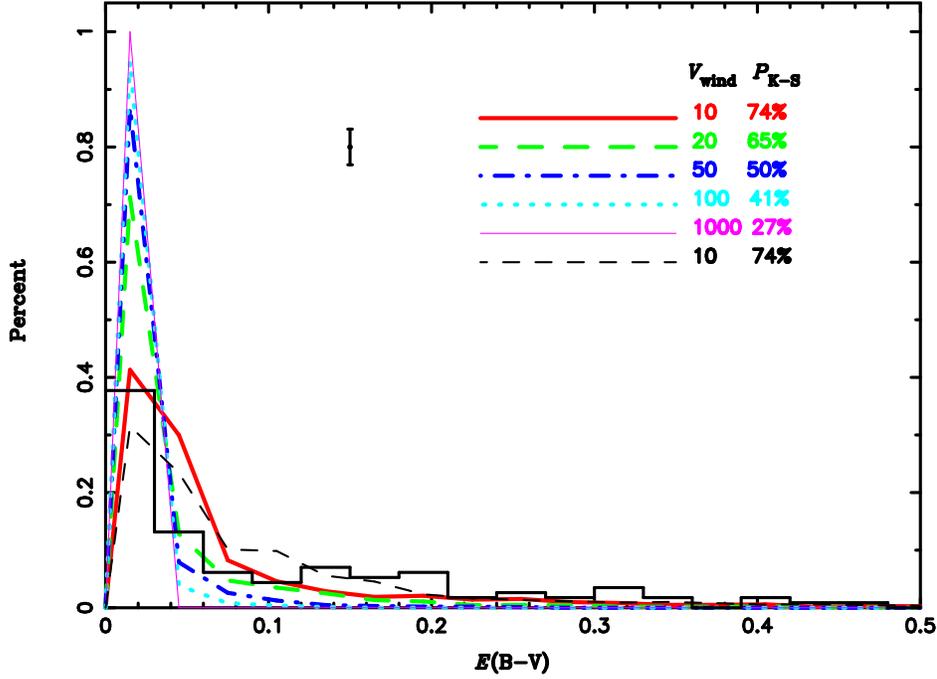}
%   \hspace{3mm}
%   \psfig {figure=mix.ps,width=80mm,height=60mm,angle=270.0}
%   \parbox{180mm}{{\vspace{2mm} }}
   \caption{The distribution of color excesses of SNe Ia at maximum light for $Z=0.02$
   and various wind velocities (in km s$^{\rm -1}$).
   The solid histogram is from observation, and the bar represents its maximum error (\citealt{REINDL05}). Via K-S test, the
   percentages
   show the confident level that the distributions of color excess from theory and observation are indistinguishable.
   The black dashed line is the best fitted line, where the weights for the components of Z=0.01, 0.02 and 0.03 are 20\%, 10\% and 70\%, respectively.}
              \label{ebvvv}%
    \end{figure}

\section{RESULTS}
\label{sect:3} In Fig. \ref{ebvzz}, we show the distribution of
the color excesses of SNe Ia at maximum light for various
metallicities. The distribution from observation is also shown by
solid histogram in the figure (\citealt{REINDL05}). We see in the
figure that the SD model with optically thick wind can reproduce
the distribution of color excesses of SNe Ia observed. The K-S
test shows that the cases of $Z=0.02$ and $Z=0.03$ have the
highest confident level that the distributions from theory and
observation are indistinguishable. The sample of \citet{REINDL05}
includes the SNe Ia with various metallicities. However, it is
difficult to determine the metallicity of the host galaxy of every
SN Ia for a large distance (\citealt{HAMUY00}). \citet{WANGXF06}
collected the properties of several SNe Ia from previous papers
and noticed that the mean metallicity of the host galaxies of the
SNe Ia is $[12+\log({\rm O/H})]_{\rm mean}=8.85\pm0.10$ (private
communication), which is consistent with solar metallicity
($[12+\log({\rm O/H})]_{\rm solar}\sim8.8$ (\citealt{ZARITSKY94}).
Then, our result is consistent with observation although the
sample collected by \citet{WANGXF06} is small.

There are some SNe Ia whose color excesses are very large (i.e.
larger than 1.0), such as SN1999cl (\citealt{JHA02};
\citealt{REINDL05}), SN2003cg (\citealt{ELIAS06}) and SN2006X
(\citealt{WANGXF07}). In the sample of \citet{REINDL05}, only one
among 113 SNe Ia has a color excess larger than 1. In our
simulation, the possibility of the high-color-excess SNe Ia is
2.0\% for $Z=0.03$, 0.4\% for $Z=0.02$ and 0 for $Z<0.02$,
consistent with that of the observations. This result might imply
that the SNe Ia may have a color excess larger than 1 only when
their host galaxies have a metallicity larger than $0.02$.
Observationally, the host galaxies of SN1999cl and SN2006X, i.e.
NGC4501 and NGC4321, are both oversolar galaxy
(\citealt{CAPUTO00}; \citealt{DORSJR06}). Although there is no
information about the metallicity of the host galaxy of SN2003cg
(NGC3169), NGC3169 is a Sa galaxy and probable has a oversolar
metallicity (\citealt{WILLNER85}).

\section{DISCUSSION}
\label{sect:4} Our analytic model is so simple. We discuss various
uncertainties about our model in this section.

\subsection{wind velocity}
\label{sect:4.1}

The major uncertainty of our model is from the assumption that
$V_{\rm wind}=10$ km s$^{\rm -1}$. Although the consistency
between the theoretical distribution of color excess and that from
observation upholds this assumption, there is no direct
observational evidence to verify it. Many observational efforts
were projected to find CSM (\citealt{HAMUY03};
\citealt{ALDERING06}; \citealt{PANAGIA06}; \citealt{OFEK07};
\citealt{PATAT07a, PATAT07b}), and only one observation obtained
the constraint of the wind velocity (\citealt{PATAT07a}). The wind
velocity is constrained to be smaller than 50 km s$^{\rm -1}$ (see
Fig. 2 in \citealt{PATAT07a}), and then Patat et al. suggested
that the progenitor of SN2006X should be a WD+RG system. However,
we can not rule out the possibility that the observed CSM is from
the optically thick wind since the companion has not been directly
observed (\citealt{HKN08}). Recently, \citet{BADENES07} explored
the relationship between the SD models with optically thick wind
for SNe Ia and the properties of the supernova remnants that
evolve after the explosion. They found that the optically thick
wind with velocity larger than 200 km s$^{\rm -1}$ would excavate
large low-density cavities around the progenitors. The large
cavities are incompatible with the dynamics of the forward shock
and the X-ray emission from the shocked ejecta in all the SNe Ia
remnants as they examined. However, they also showed that if a
wind velocity of 10 km s$^{\rm -1}$ is adopted, the properties of
type Ia supernova remnants are well compatible with the prediction
from the SD model with optically thick wind. Generally, the escape
velocity from a white dwarf is in an order of magnitude of
$10^{\rm 3}$ km s$^{\rm -1}$, which is upheld by observations from
recurrent novae (\citealt{WOOD00}). In the theoretically framework
laid down by \citet{HAC96}, an optically thick wind is formed in
CO WD envelope with photospheric velocity of $\sim10^{\rm 3}$ km
s$^{\rm -1}$ (\citealt{HAC99a, HAC99b}), where the expansion of
the photosphere is driven by helium flash in a helium shell on top
of CO WD (\citealt{KH99}). Then, in both theory and observation,
it seems not to uphold an assumption of a low wind velocity at
present. We check the influence of wind velocity on the
distribution of color excess for the case of $Z=0.02$ and the
results are shown in Fig. \ref{ebvvv}. We see in the figure that
when $V\geq 100$ km s$^{\rm -1}$, K-S test gives a low confident
level that the distributions of color excess from theory and
observation are indistinguishable, while when $V\leq 50$ km
s$^{\rm -1}$, K-S test shows an acceptable level. This result is
similar to that in \citet{BADENES07}.

\subsection{delay time}
\label{sect:4.2}

The secondary uncertainty of our model is from $t_{\rm d}$. Since
optically thick wind may stop before SN Ia explosion, $t_{\rm d}$
used in this paper overestimates the delayed time of the wind for
some systems. However, this is not a serious problem since for
most cases, SNe Ia occur during the wind phase or after the wind
phase for a short time (\citealt{HAN04}; \citealt{MENGXC09a}). In
addition, $t_{\rm d}$ obtained from equation (10) only
approximates the mean value for a given $M_{\rm 2}^{\rm i}$. This
is also not a serious problem since we only check an average
distribution of color excess. Generally, the peak of the
distribution of color excesses moves to lower color excess with
$t_{\rm d}$, and the peak value increases while the percentage of
high color-excess SNe Ia decreases with $t_{\rm d}$.

\subsection{metallicity}
\label{sect:4.3}

The combination of different metallicities is checked and shown in
Fig. \ref{ebvzz} by the black dashed line since not all SNe Ia are
from a unique metallicity. The weights of the components for
Z=0.01, 0.02 and 0.03 are simply fitted from the samples of
\citet{HAMUY00} and \citet{WANGXF06}. We also show the best fitted
line in Fig. \ref{ebvvv} by the black dashed line. We can see from
these lines that the combination of different metallicities does
not significantly increase the confident level that the
distributions for theory and observation are indistinguishable.
Actually, for any combination of $Z=0.01$, $0.02$ and $0.03$, the
results will be acceptable, i.e. yielding similar K-S test
probability, which is derived from similar K-S test probability
for the distributions of $Z=0.01$, $0.02$ and $0.03$. So, the
basic results here still hold.

\subsection{interstellar extinction}
\label{sect:4.4}

While we are not sure about the existence of the optically thick
wind (it is a prediction by a model, not an observed fact,
\citealt{HAC96}), we are absolutely sure about the existence of
the interstellar extinction (IE). In the paper, we did not
consider the influence of IE since it is difficult to separate it
from CSM dust. However, we know that Type Ia SNe, as any other
celestial object, suffer from interstellar reddening, arising in
material that has nothing to do with the circumstellar environment
of the exploding star. Let's consider the case of SN 2006X, whose
color excess is certainly larger than E(B-V)=1, while is most
likely arising in a cold molecular cloud which has nothing to do
with the explosion site (\citealt{WANGXF07}). So, for the case
suffering interstellar reddening, we should remove the influence
of IE on color excess. But at present, it is very difficult to do
this. However, it is a clear effect that IE will lead to certain
directions of where parameters such as $V_{\rm Wind}$ and $V_{\rm
ejec}$ should go, e.g. moving some very reddened observed SNe Ia
to lowering reddened one would  yield a higher $V_{\rm Wind}$ or
$V_{\rm ejec}$. For example, if we assume rather arbitrarily and
simply that all observed SNe Ia suffer a extinction of
$E(B-V)_{\rm host}=0.1$ within its host galaxy, our model
\textbf{suggests} $V_{\rm Wind}=20$ km s$^{\rm -1}$ or $V_{\rm
ejec}=20000$ km s$^{\rm -1}$. Similarly, if $E(B-V)_{\rm
host}=0.2$, our model \textbf{suggests} $V_{\rm Wind}=50$ km
s$^{\rm -1}$ or $V_{\rm ejec}=50000$ km s$^{\rm -1}$. So, the
effect of IE could be counteracted by another uncertainties,
$V_{\rm ejec}$, and the basic results in this paper are still
\textbf{valid}. (see the discussion about ejecta velocity below).

If the color excess is from CSM as shown in this paper, one may
argued that we would have seen the CSM material through radio and
X-ray emission, arising in the shock produced by the fast moving
SN ejecta crashing into the CSM (\citealt{STOCKDALE06}). However,
\citet{PANAGIA06} showed a non-detection result for radio for 27
SNe Ia via the Very Large Array (VLA) observations, even including
SN 2002ic, which indicated that mass-lose rate should be lower
than $\sim3\times10^{\rm -8}$ $M_{\odot}{\rm yr^{\rm -1}}$. We
take a typical case to check whether our model contradicts with
observations. We set companion mass to $2.1M_{\odot}$ (see Fig .10
in \citealt{MENGXC09a}), which corresponds to a delay time of
$2.5\times10^{\rm 6}{\rm yr}$. The mass of lost material is set to
$0.3M_{\odot}$ (see Fig. \ref{lostm}). The mean mass-lose rate is
$12\times10^{\rm -8}$ $M_{\odot}{\rm yr^{\rm -1}}$, which is
higher but still comparable to that inferred from observations.
Attempts to detect radio emission from SN 2002ic with the VLA were
unsuccessful (\citealt{BERGER03}; \citealt{STOCKDALE03}). SN
2002ic is the first case to show the signal of CSM
(\citealt{HAMUY03}) and the amount of the CSM is 0.5-6$M_{\odot}$
(\citealt{WAN04}; \citealt{CHU04}; \citealt{UEN04};
\citealt{KOT04}) and its properties may be explained successful by
the WD+MS model used here (\citealt{HAN06}). In addition, another
recent twins of SN 2002ic (SN 2005gj, \citealt{ALDERING06}) has
also not been detected using the VLA (\citealt{SF05}). These
non-detection results may indicate that the mechanism that is
successful used for SN Ib/c may not work for SNe Ia
(\citealt{PANAGIA06}). So, our model is at least not inconsistent
with observation at present.

\begin{figure}
   \centering
%   \vspace{2mm}
%   \begin{center}
   %%%\includegraphics{empty.eps}
   %%%\includegraphics{empty.eps}
   \includegraphics[width=90mm,angle=270.0]{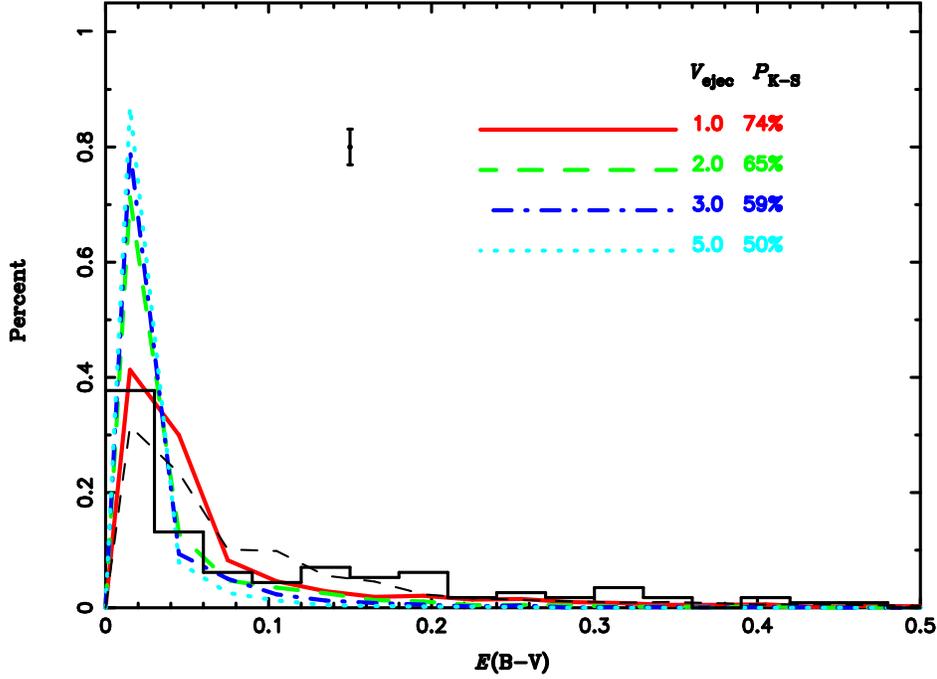}
%   \hspace{3mm}
%   \psfig {figure=mix.ps,width=80mm,height=60mm,angle=270.0}
%   \parbox{180mm}{{\vspace{2mm} }}
   \caption{The distribution of color excesses of SNe Ia at maximum light for different ejecta velocity,
   where wind velocity is assumed to be 10 km s$^{\rm -1}$. The ejecta velocity is unit in $10^{4}$ km s$^{\rm -1}$. The black dashed line is the best fitted line,
   where the weights for the components of Z=0.01, 0.02 and 0.03 are 20\%, 10\% and 70\%, respectively.}
              \label{ebvvej}%
    \end{figure}

\subsection{ionization and ejecta velocity}
\label{sect:4.5}

Generally, when a SN explodes, its radiation field is quite
strong, especially at distance $r_{\rm H}<10^{\rm 16}$ cm (for SN
2006X, $r_{\rm H}<4\times10^{\rm 15}$, \citealt{PATAT07a}). This
radiation can easily ionize hydrogen and evaporate dust up to
quite large distances, causing the disappearance of reddening. In
our simple model, this effect is not considered completely. We
check the influence of the effect.

We set $r_{\rm H}=10^{\rm 16}$ cm and assume that all hydrogen
atoms are ionized within the shell with radius of $r_{\rm H}$. The
column density of ionized hydrogen is $n_{\rm H,0}=\int_{r_{\rm
1}}^{r_{\rm H}}ar^{\rm -2}dr$, where $r_{\rm 1}=1.73\times10^{\rm
15} {\rm cm}$ (a ejecta velocity of $10^{\rm 4}$ km s$^{\rm -1}$
and a rise time of 20 day are assumed). Then, the relative
uncertainty from the ionization effect is $\frac{n_{\rm
H,0}}{n_{\rm H}}=\frac{(r_{\rm H,0}-r_{\rm 1})r_{\rm m}}{(r_{\rm
m}-r_{\rm 1})r_{\rm H,0}}\simeq\frac{r_{\rm H,0}-r_{\rm 1}}{r_{\rm
H,0}}\simeq0.83$ ($r_{\rm m}\gg r_{\rm 1}$). The uncertainty seems
too large. However, this situation may be improved by taking a
higher ejecta velocity and a smaller distance $r_{\rm H}$. A
distance of $r_{\rm H}=4\times10^{\rm 15}$ like SN 2006X may
reduce the uncertainty to 0.57.

The ejecta velocity used here may be smaller than the terminal
velocity of SN Ia which can be as high as $3\times10^{\rm 4}$ km
s$^{\rm -1}$ (\citealt{WANGLF06}). This value may reduce the
uncertainty from the ionization effect to $0\sim0.48$, which means
that no more than a half of CSM are evaporated up to large
distances and have no influence on the color excess of SNe Ia.
Then, the ionization effect at most decreases the confident level
that theory and observation are indistinguishable from 74\% to
65\%.

In Fig. \ref{ebvvej}, we show the influence of ejecta velocity on
the distribution of the color excess of SNe Ia. The influence is
similar to that of wind velocity. From the figure, we can see that
the peak of the distribution increases with the ejecta velocity,
and when $V_{\rm ejec}<5\times10^{\rm 4}$ km s$^{\rm -1}$, the
distributions of color excess from theory and observation are
indistinguishable on the confident level of higher than 50\%.
Fortunately, the effect of ionization and a higher ejecta velocity
is same to that of interstellar, and their influence on the
decrease of the confident level may counteract each other. Then,
we can say that the distribution of color excess between theory
and observation are indistinguishable on the level of higher than
59\%, even when $V_{\rm ejec}=3\times10^{\rm 4}$ km s$^{\rm -1}$.

For the discussion above, our model may get a meaning result
although our analytic model is so simple.

\section{SUMMARY and CONCLUSIONS}
\label{sect:5}

In summary, if a wind velocity of $\sim10$ km s$^{\rm -1}$ is
adopted, the SD model with optically thick wind may reproduce the
distribution of color excess of SNe Ia obtained from observation,
which might support the SD model for SNe Ia. However, if a wind
velocity larger than 100 km s$^{\rm -1}$ is adopted, the
reproduction is bad. Our results are similar to those of
\citet{BADENES07}. Then, it should be encouraged to perform more
detailed observations about the velocity of CSM around progenitors
of SNe Ia.

\normalem
\begin{acknowledgements}
This work was funded by the
 National Natural Science Foundation of China
 (NSFC) under Nos.11080922 and 12345678.
\end{acknowledgements}

\appendix                  %%appendicial material is supported

\section{This shows the use of appendix}
%A postscript file is actually an ASCII text file (you may even edit it).
%However, you need to transfer a PDF file or any compressed or packaged
%file in binary mode when using FTP.

\label{lastpage}

\end{document}